\documentclass[aps,twocolumn,pra,superscriptaddress,letterpaper,nofootinbib]{revtex4}
\usepackage{amsmath,amssymb}
\usepackage[pdftex]{graphicx}
\usepackage{hyperref}
\usepackage{mathptmx}
\usepackage{bm}

\begin{document}


\newcommand{\be}{\begin{equation}}
\newcommand{\ee}{\end{equation}}
\newcommand{\R}[1]{\textcolor{red}{#1}}
\newcommand{\B}[1]{\textcolor{blue}{#1}}


\title{Enhancing the bandwidth of gravitational-wave detectors with unstable optomechanical filters}

\author{Haixing Miao}
\affiliation{School of Physics and Astronomy,
University of Birmingham, Birmingham, B15 2TT, United Kingdom}
\author{Yiqiu Ma}
\affiliation{School of Physics, University of Western Australia, Western Australia 6009, Australia}
\author{Chunnong Zhao}
\affiliation{School of Physics, University of Western Australia, Western Australia 6009, Australia}
\author{Yanbei Chen}
\affiliation{Theoretical Astrophysics 350-17, California Institute of Technology, Pasadena,
CA 91125, USA}


\begin{abstract}
For gravitational-wave interferometric detectors, there is a tradeoff between the detector bandwidth and peak sensitivity when focusing on the shot noise level. This has to do with the frequency-dependent propagation phase lag (positive dispersion) of the signal. We consider embedding an active unstable filter---a cavity-assisted optomechanical device operating in the instability regime---inside the interferometer to compensate the phase, and using feedback control to stabilize the entire system. We show that this scheme in principle can enhance the bandwidth without sacrificing the peak sensitivity. However, there is one practical difficulty for implementing it due to the thermal fluctuation of the mechanical oscillator in the optomechanical filter, which puts a very stringent requirement on the environmental temperature and the mechanical quality factor.
\end{abstract}

\maketitle

\begin{figure}[!b]
\includegraphics[width=0.48\textwidth]{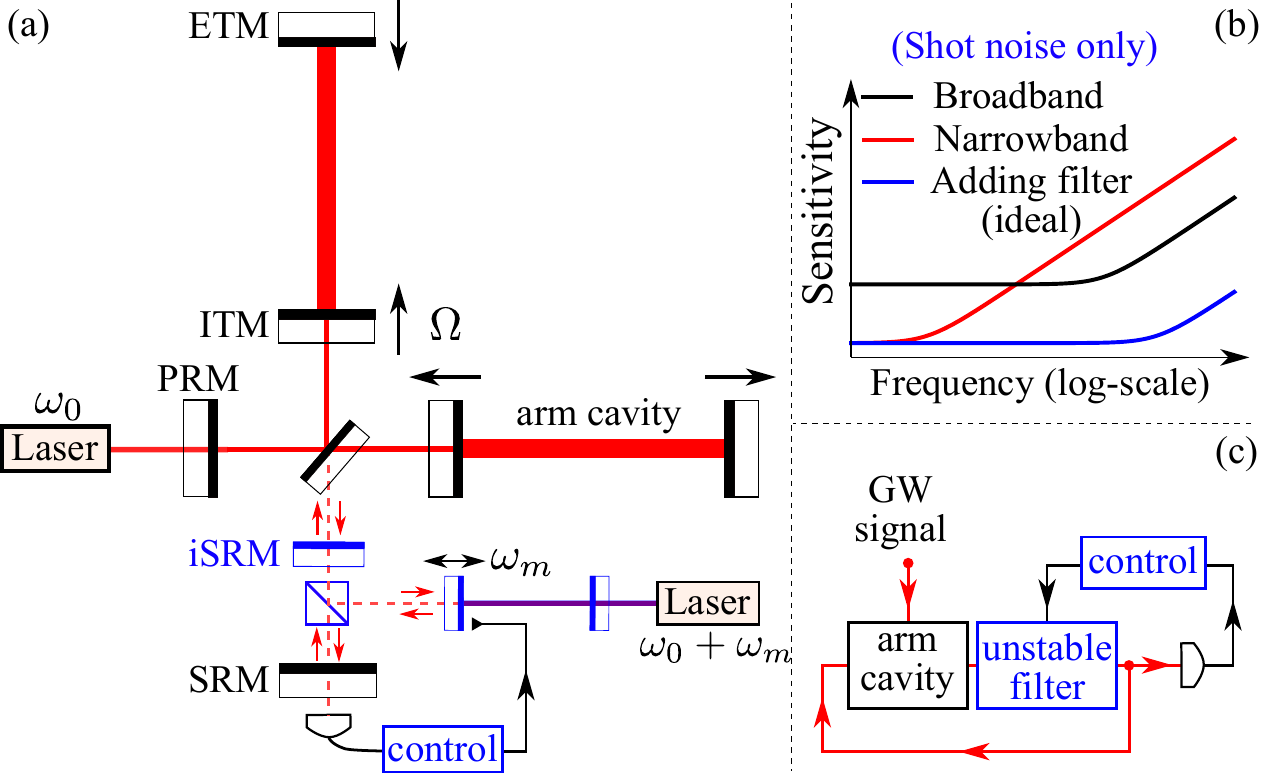}
\caption{(a) The dual-recycled Michelson configuration for advanced gravitational-wave detectors, with an unstable optomechanical filter (blue) embedded. An internal SRM (iSRM) is introduced to match the transmissibility of input test mass (ITM) mirror to remove the arm cavity pole; (b) The tradeoff between the detector bandwidth and peak sensitivity (black and red curves), and the effect of adding unstable filter in the ideal scenario (blue); (c) A simple flow chart.
\label{fig:configuration}}
\end{figure}

{\it Introductions.---}Advanced gravitational-wave (GW) detectors, including Advanced LIGO\,\cite{Gregg:aLIGO2010}, Advanced VIRGO\,\cite{aVir2009} and KAGRA\,\cite{Somiya2012a}, are dual-recycled Michelson interferometers with both power-recycling mirror (PRM) and signal-recycling mirror (SRM), as shown schematically in Fig.\,\ref{fig:configuration}(a). The GW-induced differential motion of the two end test masses (ETMs) modulates the laser at frequency $\omega_0$ and creates sidebands at $\omega_0\pm\Omega$ with $\Omega$ being the GW frequency, ranging from 10Hz to $10^4$Hz for ground-based detectors. In the presence of SRM, these sidebands are coherently reflected back to the interferometer, which modifies the detector frequency response and bandwidth. For instance, advanced LIGO, in its nominal operation mode, is using SRM to broaden the detector bandwidth. However, this is at a price of decreasing the peak sensitivity when considering the shot noise, as illustrated in Fig.\,\ref{fig:configuration}(b). Such a tradeoff between the bandwidth and the peak sensitivity, known as the Mizuno theorem\,\cite{Mizuno1995}, can be attributable to the frequency-dependent propagation phase of sidebands (positive dispersion), as cavity resonant condition is satisfied only for a single frequency. There are several approaches proposed in the literature aiming at broadening the detector bandwidth without degrading its peak sensitivity. One is the so-called {\it white-light-cavity} idea\,\cite{Wicht1997, Wicht2000, Wise2004, Wise2005, Pati2007,Yum2013, Zhou2014, Ma2015} that uses atomic medium with negative dispersion to cancel the positive dispersion. The other is applying either external squeezing\,\cite{Cav1980, Kirk:2004} or internal squeezing\,\cite{Rehbein2005, Peano2015} to reduce the shot-noise level while keeping a broad detector bandwidth. In particular, external squeezing has already been implemented in large-scale GW detectors\,\cite{GEO:Squeezing,Aasi2013} and also planned for future upgrades\,\cite{Evans2013, Miao2014a,ISWP2014}.

We consider a different approach to realize such a broadband enhancement by placing an optomechanical device as a filter inside the signal-recycling cavity, illustrated schematically in Fig.\,\ref{fig:configuration}(a). Such an optomechanical filter operates in the unstable regime pumped by an additional laser, and the entire scheme is stabilized via feedback control. Within the signal loop, the unstable filter acts as a phase compensator with negative dispersion. Notice that there is no violation of causality, as the controlled system is stable and the measured output always lags behind incoming GW signals. In addition, the feedback uses the SRM output, as indicated by the flow chart in Fig.\,\ref{fig:configuration}(c), which contains the GW signal together with noise, and therefore it does not influence the signal-to-noise ratio, as proven in Refs.\,\cite{BuonannoChen2002, Harris2013}. This idea is inspired by recent experimental observations of the optomechanical analogue of the electromagnetically induced transparency by S. Weis {\it et al.}\,\cite{Weis2010}, Teufel {\it et al.}\,\cite{Teufel2011a} and Safavi-Naeini {\it et al.}\,\cite{Safavi-Naeini2011}, and also a more recent theoretical proposal by Ma {\it et al.} \,\cite{Ma2014a}. These are, instead, using the optomechanical interaction to enhance the phase lag within a certain frequency range.


{\it A brief summary.---}Before develling into the details, we summarize the main features and explain them qualitatively. For the optomechanical filter, the radiation pressure couples the optical field intensity and the mechanical displacement. Such a nonlinear coupling is analogous to the three-wave mixing in nonlinear optics. By tuning the pump laser frequency to be $\omega_0+\omega_m$ with the mechanical oscillator frequency $\omega_m$ much larger than the cavity bandwidth $\gamma_f$, i.e.,
\begin{equation}\label{eq:resolved_sideband}
\omega_m \gg \gamma_f \gg \Omega\,,
\end{equation}
the cavity resonance at $\omega_0$ is in favor of the down-conversion process, which amplifies the sidebands around $\omega_0$ and mechanical motion. It can be viewed as a phase-insensitive parametric amplifier for sidebands with the following input-output relation:
\begin{equation}\label{eq:phase_insensitive_amplifier}
\hat a_{\rm out}(\Omega)\approx \frac{\Omega+i(\gamma_m+\gamma_{\rm opt})}{\Omega+i(\gamma_{m}-\gamma_{\rm opt})}\hat a_{\rm in}(\Omega)\,.
\end{equation}
Here $\gamma_m\equiv \omega_m/Q_m$ is the mechanical damping rate with $Q_m$ being the quality factor; $\gamma_{\rm opt}$ is the negative mechanical damping rate due to the optomechanical interaction, and is approximately equal to $P_c {\cal F} \omega_0/(m \omega_m c^2)$, where $P_c$ is the intra-cavity laser power, $\cal F$ is the cavity finesse, and $m$ is the mass of the mechanical oscillator.

\begin{figure}[!t]
\includegraphics[width=0.45\textwidth]{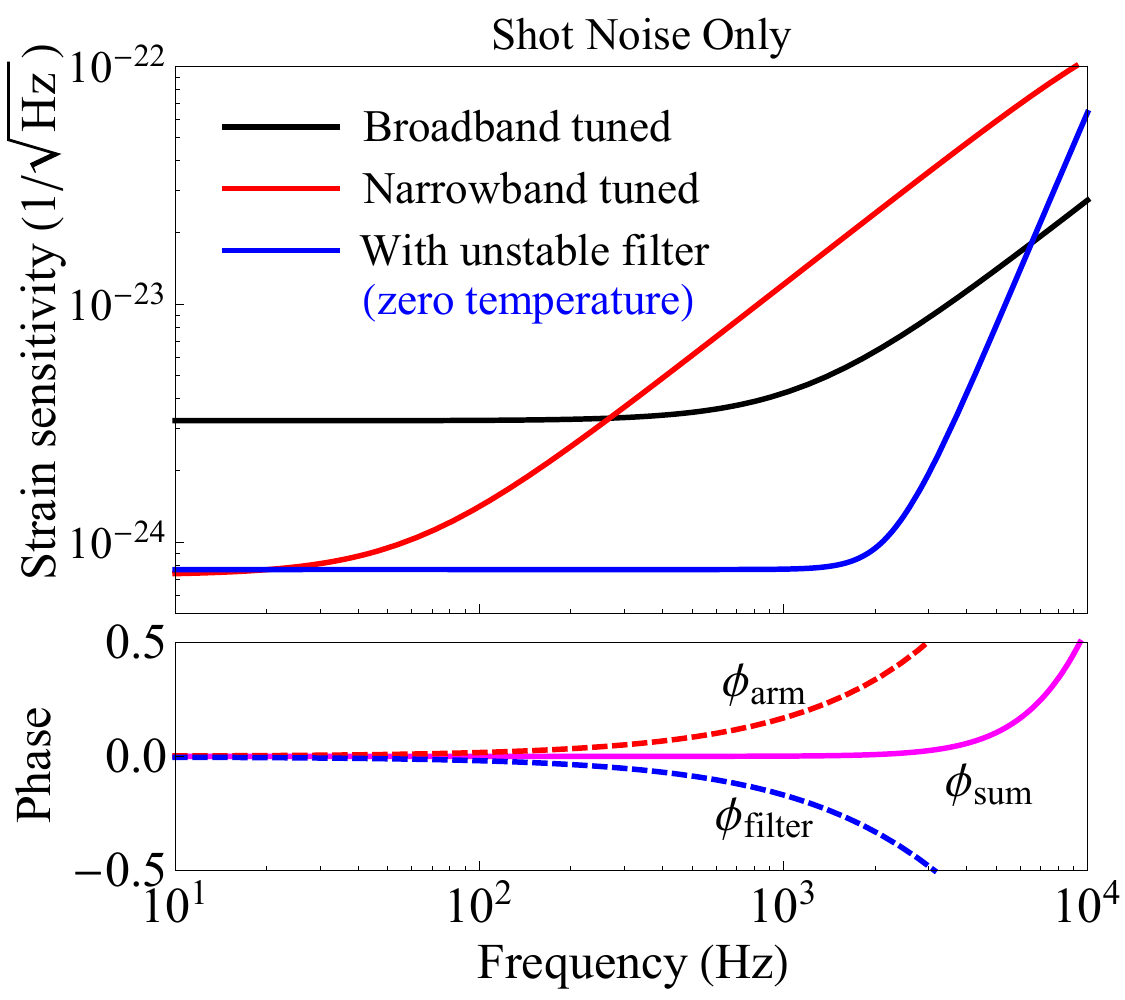}
\caption{The top figure shows the shot-noise-only strain sensitivity of the unstable-filter scheme, compared with the broadband and narrowband tuned cases without the filter.
The power and arm cavity length are the same as Advanced LIGO. At high frequencies, its sensitivity deviates from the ideal scenario because of imperfect phase cancellation of higher-order terms of $\Omega$ in $e^{i\phi_{\rm arm}(\Omega)}$ (bottom figure).
\label{fig:sensitivity}}
\end{figure}

The unstable regime we referred to is when $\gamma_{\rm opt}$ becomes much larger than $\gamma_m$ and the mechanical damping rate becomes negative. With the feedback control engaged, the open-loop input-output relation of the optomechanical filter is
\begin{equation}\label{eq:phase_insensitive_amplifier_approximation}
\hat a_{\rm out}(\Omega)\approx \frac{\Omega+i\gamma_{\rm opt}}{\Omega-i\gamma_{\rm opt}}\hat a_{\rm in}(\Omega)\approx -\exp\left(-\frac{2i\Omega}{\gamma_{\rm opt}}\right)\hat a_{\rm in}(\Omega)\,,
\end{equation}
which exhibits negative dispersion rather than positive dispersion for a passive filter cavity. To compensate the propagation phase lag $\phi_{\rm arm}=2\Omega L_{\rm arm}/c$ with $L_{\rm arm}$ being the arm cavity length and equal to 4km for LIGO, we therefore require
\begin{equation}\label{eq:phase_cancelation_condition}
 \gamma_{\rm opt}=c/L_{\rm arm}\,,
\end{equation}
and the resulting intra-cavity laser power scales as:
\begin{equation}\label{eq:gamma_opt_scalings}
P_c\approx 10^2 W\left(\frac{4{\rm km}}{L_{\rm arm}}\right) \left(\frac{10{\rm MHz}}{\omega_m/2\pi}\right)\left(\frac{0.1{\rm mg}}{m}\right)\left(\frac{\cal F}{10^5}\right)\,.
\end{equation}
In Fig.\,\ref{fig:sensitivity}, we show the resulting shot-noise-only sensitivity with nominal parameters given above and one additional parameter $L_f=1$cm being the filter cavity length. We can see that the unstable filter can compensate the phase at low frequencies but not perfect at high frequencies due to high-order frequency dependence of the phase delay, which in principle can be improved by cascading several unstable filters.

\begin{figure}[!b]
\includegraphics[width=0.45\textwidth]{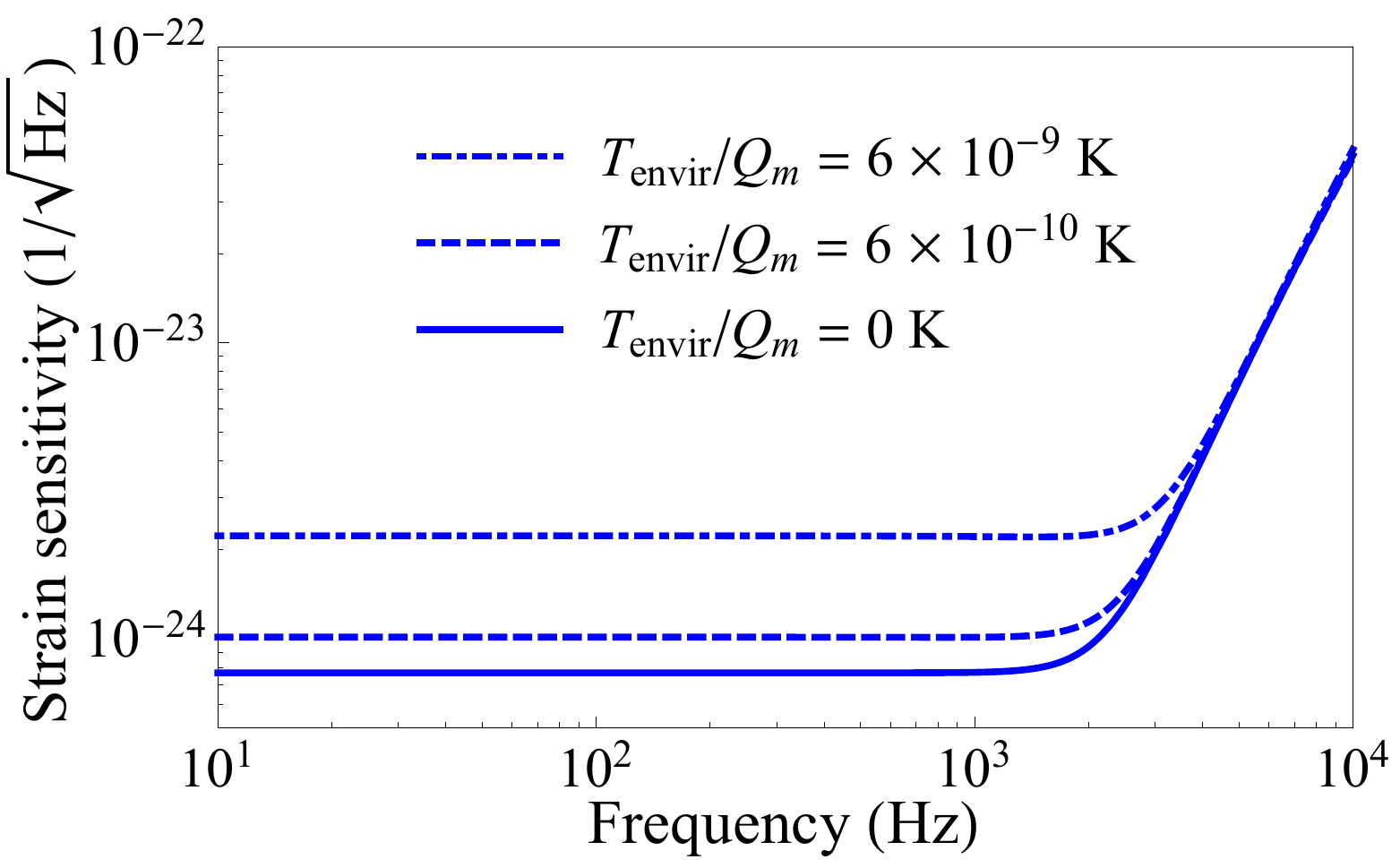}
\caption{Effect of the thermal fluctuation of the mechanical oscillator on the sensitivity.
\label{fig:temperature}}
\end{figure}

So far, we have only mentioned the phase property of the unstable filter. As a general principle proven by Caves for phase-insensitivity parametric amplifiers\,\cite{Cav1982}, there will be an additional noise term given Eq.\,\eqref{eq:phase_insensitive_amplifier}:
\begin{equation}\label{eq:additional_noise}
\hat n_{\rm add}(\Omega)=\frac{2\sqrt{\gamma_m\gamma_{\rm opt}}}{\Omega+i(\gamma_m-\gamma_{\rm opt})}\hat b^{\dag}_{\rm th}(-\Omega)\,.
\end{equation}
As shown later, it comes from the coupling of the mechanical oscillator to the environmental thermal bath. The spectral density for $\hat b_{\rm th}$ is approximately given by $2k_B T_{\rm envir}/(\hbar\omega_m)+1$ with $T_{\rm envir}$ being the temperature. In order for the thermal noise to be lower than the quantum shot noise, we require
\begin{equation}\label{eq:thermal_requirement}
8 k_B T_{\rm envir} /Q_m \lesssim{\hbar \gamma_{\rm SRM}}\,,
\end{equation}
with $\gamma_{\rm SRM}\equiv c\,T_{\rm SRM}/(4L_{\rm arm})$ and $T_{\rm SRM}$ being the power transmissivity of the SRM. Here $\gamma_{\rm SRM}$ is the original detector bandwidth, as ITM and iSRM are impedance matched and the bandwidth is solely determined by SRM. Therefore, the lower is the detector bandwidth that we start off, the higher requirement will be imposed on the temperature and quality factor.

As an order of magnitude estimation, we have
\begin{equation}\label{eq:temperature}
\frac{T_{\rm envir}}{Q_m}\lesssim6 \times 10^{-10}{\rm K} \left(\frac{\gamma_{\rm SRM}/2\pi}{100\rm Hz}\right)\,.
\end{equation}
In Fig.\,\ref{fig:temperature}, we show the thermal noise effect on the sensitivity, which is significant. As mentioned in Ref.\,\cite{Ma2014a}, one approach for mitigation is applying the optical-dilution idea\,\cite{Corbitt2007,Chang:2012fj, Ni:2012qy, Korth2013}, which uses optically-induced rigidity to dilute the mechanical dissipation, and allows for enhancement of $Q_m$ by a factor of hundred or even more but further experimental study is necessary. After this summary, we will present more details about the scheme and also the issue of feedback control.


\begin{figure}[!t]
\includegraphics[width=0.42\textwidth]{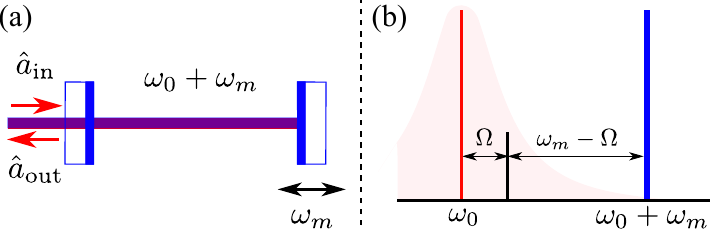}
\caption{(a) schematics of the optomechanical filter; (b) frequency of interests: cavity resonance $\omega_0$ (red), laser frequency $\omega_0+\omega_m$ (blue) and its lower sideband (black), with the upper sideband around $\omega_0+2\omega_m$ not shown.
\label{fig:frequency_map}}
\end{figure}

{\it Dynamics of the optomechanical filter.}---We start with the optomechanical filter shown in Fig.\,\ref{fig:frequency_map}(a), derive the input-output relation for the sidebands, and later combine it with the main interferometer. It is an optomechanical device which has been studied extensively in the literature (see recent reviews: Refs.\,\cite{Aspelmeyer2013, Yanbei:Review}), with Hamiltonian given by $\hat H_{\rm filter} =\hat H_{0}+\hat H_{\rm int}+\hat H_{\gamma_f}+\hat H_{\gamma_m}$. The free part $\hat H_0$ is
\begin{equation}\label{eq:free_Hamiltonian}
\hat H_0 =\hbar \omega_0 \hat a^{\dag}\hat a+\frac{\hat p^2}{2m}+\frac{1}{2}m\omega_m^2\hat x^2\,;
\end{equation}
The linearized interaction Hamiltonian $\hat H_{\rm int}$ is
\begin{equation}\label{eq:int_Hamiltonian}
\hat H_{\rm int}=-\hbar g_0 \left[\hat a \, e^{i(\omega_0+\omega_m) t}+\hat a^{\dag}e^{-i(\omega_0+\omega_m)t}\right]\hat x
\end{equation}
with $g_0\equiv \omega_0 \bar a/L_f$ and $\bar a=(2P_c L_f/(\hbar\omega_0 c))^{1/2}$; $\hat H_{\gamma_f}$ describes how the cavity mode $\hat a$ interacts with ingoing (outgoing) field $\hat a_{\rm in}$ ($\hat a_{\rm out}$); $\hat H_{\gamma_m}$ describes the coupling between the mechanical oscillator with the environmental thermal bath.

The parameter regime we are interested in, as mentioned earlier in Eq.\,\eqref{eq:resolved_sideband}, is the so-called resolved-sideband regime illustrated in Fig.\,\ref{fig:frequency_map}(b). This allows us to ignore the upper mechanical sideband around $\omega_0+2\omega_m$ and use the rotating-wave approximation (RWA) in the interaction picture, obtaining
\begin{equation}\label{eq:interaction_Hamiltonian}
\hat H_{\rm int}^{\rm RWA}=-\hbar g (\hat a\,\hat b+\hat a^{\dag}\,\hat b^{\dag})\,,
\end{equation}
where we have introduced the annihilation operator $\hat b$ for the mechanical oscillator through $\hat x(t)\equiv x_q(\hat b \, e^{-i\omega_m t}+\hat b^{\dag}e^{i\omega_m t})$ and $g\equiv g_0 x_q$ with $x_q$ being the ground-state uncertainty.

\begin{figure}[!t]
\includegraphics[width=0.42\textwidth]{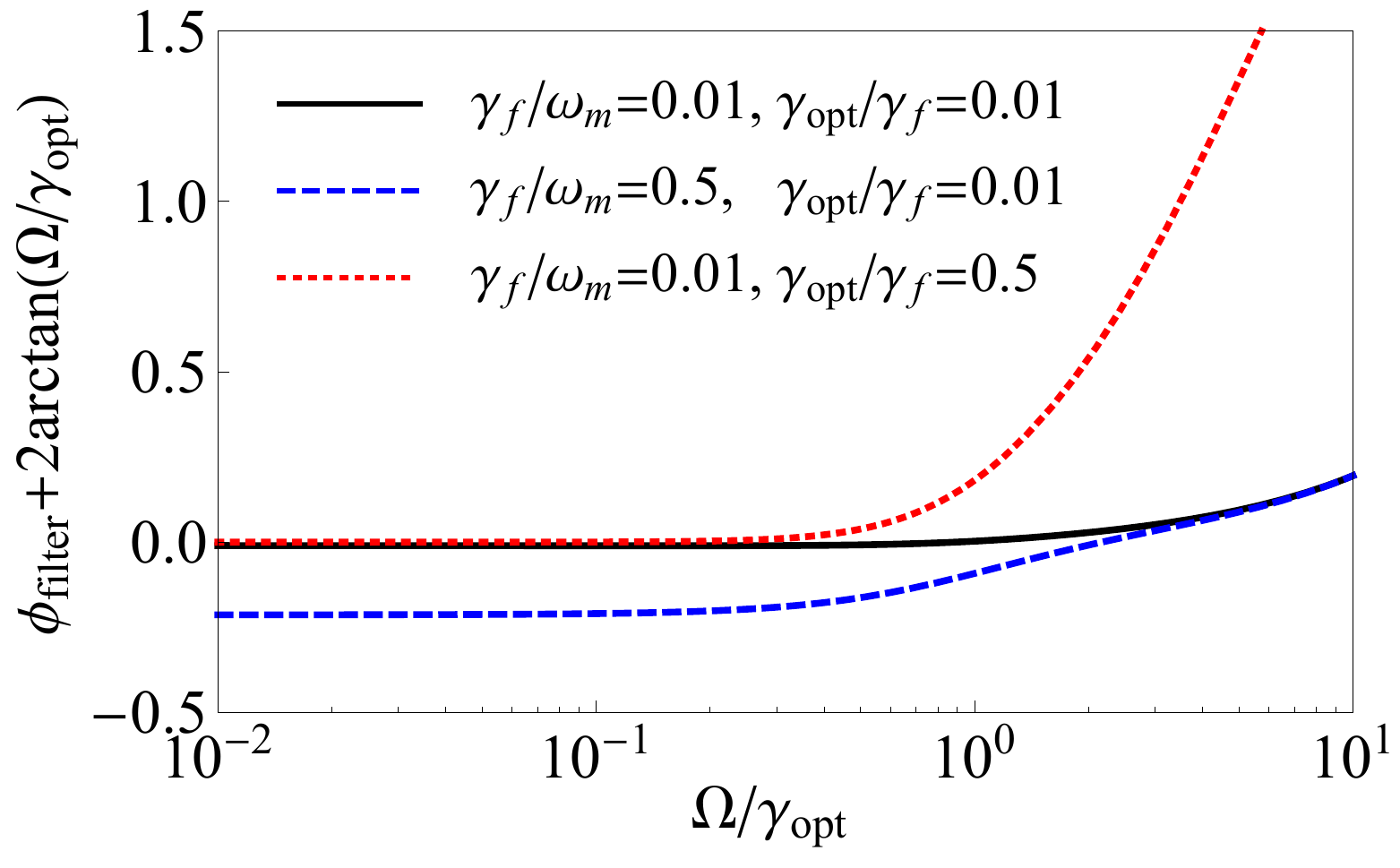}
\caption{The difference between $\phi_{\rm filter}$ obtained without using RWA and $-2\arctan(\Omega/\gamma_{\rm opt})$. The parameter regime with $\omega_m\gg\gamma_f\gg \gamma_{\rm opt}$ is thus preferred for matching the required frequency dependence.
\label{fig:phase}}
\end{figure}

The resulting Heisenberg equations of motion read:
\begin{align}\label{eq:EOM}
\dot {\hat a}(t)+\gamma_f \hat a(t)&=i g \hat b^{\dag}(t)+\sqrt{2\gamma_f}\,\hat a_{\rm in}(t)\,,\\
\dot {\hat b}(t)+\gamma_m \hat b(t)&=i g \hat a^{\dag}(t)+\sqrt{2\gamma_m}\,\hat b_{\rm th}(t)\,.
\end{align}
Solving them in the frequency domain, we obtain the input-output relation for sidebands using $\hat a_{\rm out}=-\hat a_{\rm in}+\sqrt{2\gamma_f}\,\hat a$:
\begin{equation}\label{eq:input_output_relation}
 \hat a_{\rm out}(\Omega)\approx \frac{\Omega+i(\gamma_m+\gamma_{\rm opt})}{\Omega+i(\gamma_m-\gamma_{\rm opt})}\hat a_{\rm in}(\Omega)+\hat n_{\rm add}(\Omega)\,,
\end{equation}
where we have used $\gamma_f\gg \Omega$ and defined $\gamma_{\rm opt}\equiv g^2/\gamma_f\approx P_c {\cal F} \omega_0/(m \omega_m c^2)$. The first term gives Eq.\,\eqref{eq:phase_insensitive_amplifier} and in Fig.\,\ref{fig:phase}, we compare it with the exact phase of the optomechanical filter without RWA and $\gamma_f\gg \Omega$. The second term $\hat n_{\rm add}$ is the thermal noise term mentioned earlier.

For the mechanical sideband, the resulting susceptibility is:
\begin{equation}\label{eq:chi_for_b}
\chi_m(\Omega)=-(i\Omega+\gamma_m-\gamma_{\rm opt})^{-1}\,.
\end{equation}
The mechanical motion is unstable in the parameter regime of interest with $\gamma_{\rm opt}$ much larger than the bare mechanical damping rate $\gamma_m$, and a feedback control is thus needed. We show the control scheme after considering the filter together with the main interferometer, as given below.

{\it Dynamics of the entire system.---}The total Hamiltonian  reads $\hat H_{\rm tot}=\hat H_{\rm ifo}+\hat H_{\rm filter}+\hat H_{\rm ifo-filter}$. We can model the main interferometer also as an optomechanical device\,\cite{scaling_law}:
\begin{equation}\label{eq:H_ifo}
\hat H_{\rm ifo}=\hbar \omega_0\hat d^{\dag}\hat d+\hat H_{\gamma_{\rm ifo}}+\frac{\hat P^2}{2M}-\hbar G_0(\hat d+\hat d^{\dag})\hat X+\hat X\,F_{\rm GW} \,.
\end{equation}
Here $\hat d$ is the differential optical mode---{\it a single-mode approximation} which is valid when considering sideband frequency much lower than one free spectral range $c/(2L_{\rm arm})$; $\hat H_{\gamma_{\rm ifo}}$ describes the interaction between $\hat d$ and the ingoing (outgoing) field $\hat d_{\rm in} (\hat d_{\rm out})$ at the dark port; $G_0=\omega_0\bar d/L_{\rm arm}$ is the coupling strength and $\bar d=(2P_{\rm arm} L_{\rm arm}/(\hbar\omega_0 c))^{1/2}$ with $P_{\rm arm}$ being the arm cavity power; $\hat X$ is the differential motion of the ETMs and is driven by the GW as a tidal force: $F_{\rm GW}=ML_{\rm arm}\ddot h(t)$ with $h$ being the GW strain. The coupling between $\hat d$ and $\hat a$ can be quantified by (exchanging photons):
\begin{equation}\label{eq:H_ifo-filter}
\hat H_{\rm ifo-filter}=\hbar \omega_s (\hat d^{\dag}\hat a+\hat d\,\hat a^{\dag})\,,
\end{equation}
where $\omega_s\equiv \sqrt{c\gamma_f/L_{\rm arm}}$ is the coupling rate, and equal to the optomechanical coupling rate $g$ when Eq.\,\eqref{eq:phase_cancelation_condition} is satisfied.

To focus on the shot noise, we will first ignore the radiation pressure effect on ETMs by assuming $M\rightarrow \infty$ (finite test mass is considered in the rigorous treatment). The resulting equations of motion under RWA are:
\begin{align}\label{eq:eom_total}
\dot {\hat a}&= -i\omega_s \hat d + i g \hat b^{\dag}\,, \\
\dot {\hat b}^{\dag} & = -\gamma_m \hat b^{\dag}-i g \hat a +\sqrt{2\gamma_{m}}\,\hat b^{\dag}_{\rm th}\,,\\
\dot {\hat d}&= -\gamma_{\rm SRM}\hat d-i\omega_s \hat a + \sqrt{2\gamma_{\rm SRM}}\,\hat d_{\rm in}+iG_0 L_{\rm arm}h\,.
\end{align}
The input-output relation at the SRM is $
\hat d_{\rm out}=\hat d_{\rm in}-\sqrt{2\gamma_{\rm SRM}}\,\hat d$.

The system stability can be examined from the eigenvalues of the dynamical matrix (read off from equations of motion):
\begin{equation}\label{eq:dynamical_matrix}
{\bf A}=\left[\begin{array}{ccc}
0& i g & -i\omega_s \\
-ig & -\gamma_m & 0\\
-i\omega_s & 0 & -\gamma_{\rm SRM}
\end{array}\right]\,.
\end{equation}
Having eigenvalues with positive real part implies instability, which is the case given the relevant parameter regime.

To find the stabilizing controller, we follow the state-space approach. We first need examining observability and controllability of the system. It turns out that if we detect the output phase quadrature, which contains the GW signal, using homodyne detection, only one of the two mechanical quadratures will be observed, and this will lead to an uncontrollable system. An apparent solution is measuring both the amplitude and phase quadratures with heterodyne detection. However, this is at a price of increasing the shot noise by $\sqrt{2}$ in amplitude\,\cite{BCM2003}. Instead, we can pick off a small portion of the output signal and use the heterodyne detection only for control purpose. The rest is still measured using the homodyne detection for exacting the GW signal, which will not degrade the sensitivity by an noticeable amount.

The resulting readout vector with heterodyne detection in the sideband picture is ${\bf D}=(0,\,0,\,1)$---the measured $\hat d_{\rm out}$ is linear to $\hat d$, and the control input vector is ${\bf B}=(0,\,1,\,0)^{\rm T}$ (superscript $^{\rm T}$ for transpose)---the feedback force is coupled to the mechanical displacement that is linear to $\hat b^{\dag}$. The system becomes both observable and controllable, as we have ${\rm rank}([{\bf D}; {\bf D}{\bf A}; {\bf D}{\bf A}^{2}])={\rm rank}([{\bf B}, {\bf A}{\bf B}, {\bf A}^2 {\bf B}])=3$. A stabilizing controller can then be constructed (see section 9.4 of Ref.\,\cite{Antsaklis_Michel_book}), which has the following transfer function (from $\hat d$ to $\hat b^{\dag}$) in the frequency domain:
\begin{equation}\label{eq:state_observer_feedback}
C(\Omega)=-{\bf K}(-i\Omega {\bf I}-{\bf A}+{\bf B}{\bf K}+{\bf L}{\bf D})^{-1}{\bf L},
\end{equation}
where ${\bf K}=(K_1,K_2, K_3)$ and ${\bf L} =(L_1, L_2, L_3)^{\rm T}$ are chosen such that eigenvalues of ${\bf A}-{\bf L}{\bf D}$ and ${\bf A}-{\bf B}{\bf K}$ all have negative real part. Given the nominal parameter specification in Eq.\,\eqref{eq:gamma_opt_scalings}, the system will be stabilized by setting ${\bf K}=3\times 10^5 \epsilon^{-1}(- i,\, 1,\, -1)$ and ${\bf L}=5\times 10^5 \epsilon^{-1}(i,\,1.2,\,1)$, where $\epsilon$ is the fraction of the output (in amplitude) measured using heterodyne detection. With this set of $\bf K$ and $\bf L$ found, arbitrary stabilizing controller can be generated via Youla-Ku\v{c}era parameterization\,\cite{Kucera2013}, which is used for control optimization.

\begin{figure}[!b]
\includegraphics[width=0.45\textwidth]{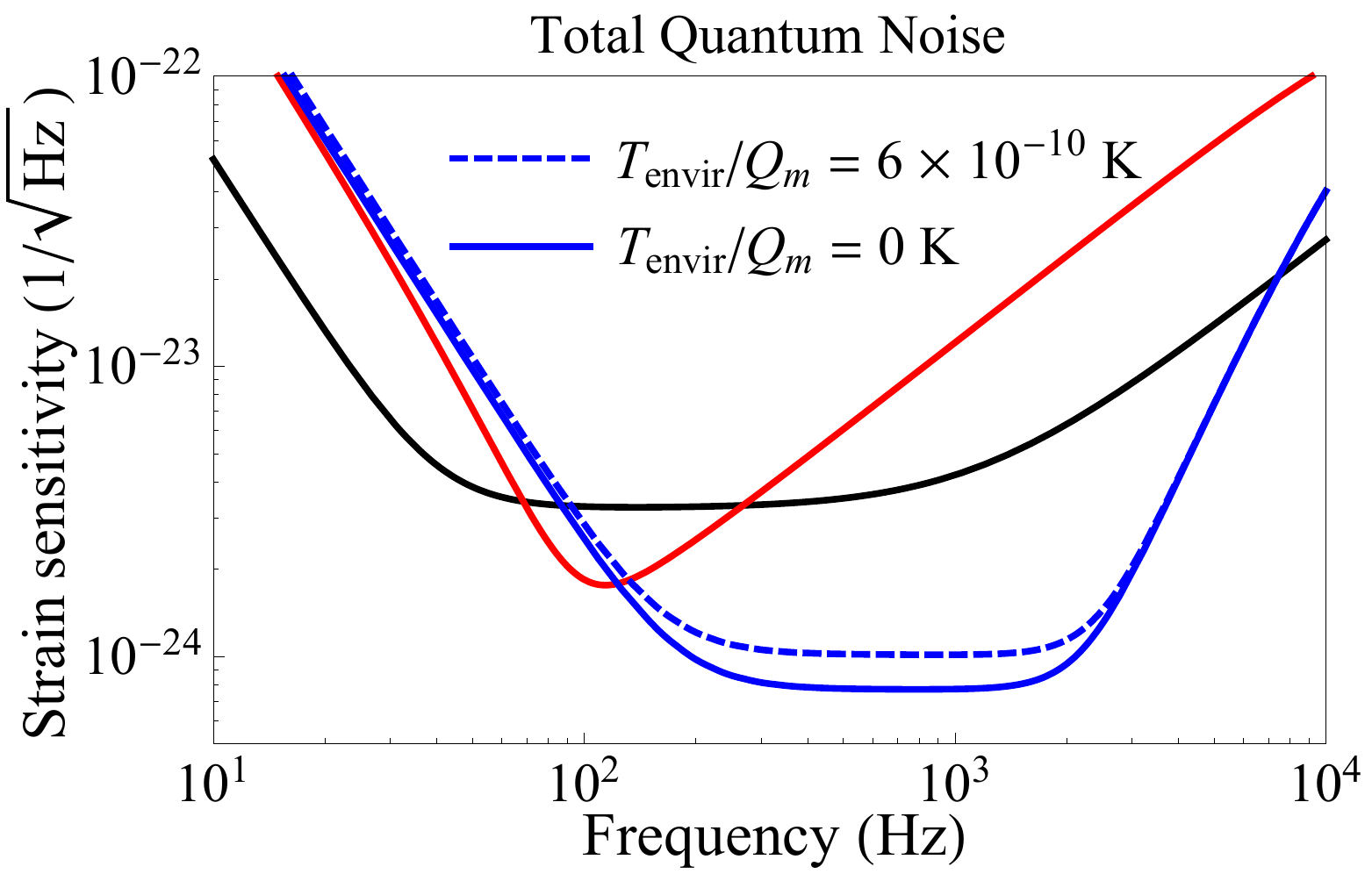}
\caption{The figure shows the total quantum noise (including radiation pressure noise) of the unstable filter scheme at different temperature. The color code for different curves is the same as Fig.\,\ref{fig:sensitivity}. The test mass is assumed to be 40kg, the same as Advanced LIGO.
\label{fig:sensitivity_new}}
\end{figure}

{\it Rigorous treatment.}---We have used several approximations in order to gain intuitive understanding. These approximations are reasonable in the parameter regime that we are focusing on. We briefly outline how the more rigorous treatment is applied in the actual analysis for producing the results presented in this paper. Specifically, the optomechanical interaction in the main interferometer mixes the upper and lower sidebands around $\omega_0$, which was ignored by assuming infinite test mass and focusing on the shot noise only. The interaction in the optomechanical filter mixes in sidebands around $\omega_0+2\omega_m$ that were ignored in RWA. A rigorous treatment therefore involves propagating four sidebands at frequency $\omega_0\pm \Omega$ and $\omega_0+2\omega_m\pm\Omega$. The result is shown in Fig.\,\ref{fig:sensitivity_new} as one example. Notice that the low-frequency radiation pressure noise can be reduced by either using frequency-dependent readout\,\cite{klmtv} or increasing the test mass size.

{\it Conclusions.---}We have considered using an unstable optomechanical filter to enhance the bandwidth of advanced gravitational-wave detectors. The entire system is stabilized via a feedback control. Within the loop, phase lag of the sidebands due to free propagation inside the arm cavity is compensated by the unstable filter. The issue for implementing this scheme is the stringent requirement on thermal noise of the mechanical oscillator in the filter, and could be mitigated by using the optical-dilution idea but requires further experimental verification.


{\it Acknowledgements.---}We thank H. Yang, M. Wang and other members of the LIGO-MQM discussion group for
fruitful discussions. H.M. is supported by the Marie-Curie Fellowship. Y.C. is supported by NSF Grants PHY-1068881 and CAREER Grant PHY-0956189. Y.M. and C.Z. has been supported by the Australian Research Council.


\bibliography{references}


\end{document}